\documentclass[%
 amsmath,amssymb,
 aps,
prb,
]{revtex4-1}
\usepackage{blindtext}
\usepackage{graphicx}
\usepackage{dcolumn}
\usepackage{bm}
\usepackage{hyperref}
\usepackage{color}
\usepackage{braket}
\usepackage{ulem}
\begin{document}


\title{First-principles study of electronic transport and structural properties of Cu$_{12}$Sb$_4$S$_{13}$ in its high-temperature phase}

\author{Cono Di Paola}
 \email{cono.di\_paola@kcl.ac.uk}
 \author{Francesco Macheda}
 \email{francesco.macheda@kcl.ac.uk}
  \author{Savio Laricchia}
 \email{savio.laricchia@kcl.ac.uk}
 \author{Cedric Weber}
 \email{cedric.weber@kcl.ac.uk}
\author{Nicola Bonini}%
 \email{nicola.bonini@kcl.ac.uk}
\affiliation{%
Department of Physics, King's College London, Strand, London WC2R 2LS, United Kingdom
}%

\date{\today}

\begin{abstract}
We present an {\it ab initio} study of the structural and electronic transport properties of tetrahedrite, Cu$_{12}$Sb$_4$S$_{13}$, in its high-temperature phase. We show how this complex compound can be seen as the outcome of an ordered arrangement of S-vacancies in a semiconducting fematinite-like structure (Cu$_3$SbS$_4$).
Our calculations confirm that the S-vacancies are the natural doping mechanism in this thermoelectric compound and
reveal a similar local chemical environment around crystallographically inequivalent Cu atoms, shedding light on the debate on XPS measurements in this compound. 
To access the electrical transport properties as a function of temperature we use the Kubo-Greenwood formula applied to snapshots of first-principles molecular dynamics simulations.
This approach is essential to effectively account for the interaction between electrons and lattice vibrations in such a complex crystal structure where a strong anharmonicity plays a key role in stabilising the high-temperature phase.
Our results show that the Seebeck coefficient is in good agreement with experiments and the phonon-limited electrical resistivity displays a temperature trend that compares well with a wide range of experimental data. The predicted lower bound for the resistivity turns out to be remarkably low for a pristine mineral in the Cu-Sb-S system but not too far from the lowest experimental data reported in literature.
The Lorenz number turns out to be substantially lower than what expected from the free-electron value in the Wiedemann-Franz law, thus providing an accurate way to estimate the electronic and lattice contributions to the thermal conductivity in experiments, of great significance in this very low thermal conductivity crystalline material.

\end{abstract}

\maketitle

\section{Introduction}
Within the family of copper-based semiconductors, the ternary Cu-Sb-S system has attracted great interest in recent years due to a variety of appealing structural, electronic and thermal transport properties. Indeed, Cu-Sb-S compounds display a rich structural variety, a large range of band gaps and are characterised by extremely low thermal conductivities. These features combined with the non-toxicity and abundance of the constituent elements make the Cu-Sb-S system an ideal playground to explore and optimise materials for sustainable thermoelectric devices.

Within this system, the tetrahedrite Cu$_{12}$Sb$_4$S$_{13}$ is a compound of particular interest as in its pristine form it has a remarkably high zT value, approximately 0.6 at 700 K, that is the result of a very low thermal conductivity (below 1 Wm$^{-1}$K$^{-1}$ from 300 to 700 K) and a high power factor~\cite{doi:10.1002/aenm.201200650}. 

In spite of a large effort in characterising the material (see, for instance, Refs.~\onlinecite{doi:10.1002/aenm.201200650,Nasonova2016,Suekuni_2013,doi:10.1021/cm404026k,CHETTY2015266}), a detailed understanding of the origin of these transport characteristics is still missing---possibly preventing further optimisations of the compound as well as the development of design ideas for novel thermoelectric materials.
For instance, the relative electronic and lattice contributions to such a low thermal conductivity are still under debate. In addition, it is not clear how the carrier scattering mechanisms determine the electronic transport and especially the temperature dependence of the electrical resistivity that appears to vary considerably in the available experimental data. In this, of particular interest is the effect on the electronic carriers of a very peculiar lattice dynamics, that shows soft phonons, strong anharmonicity and unusually large anisotropic atomic displacement parameters as shown in diffraction studies~\cite{doi:10.1002/adfm.201500766}. 

In order to address these important issues we have initially carried out an extensive study of the structural stability of the high-temperature phase of tetrahedrite. In particular, we have done this by analysing the structural relationship between tetrahedrite and the simpler fematinite structure (Cu$_3$SbS$_4$), which under optimal doping shows as well quite good thermoelectric properties~\cite{doi:10.1021/acs.jpcc.6b09379, C8TC02481B}. As we will show, tetrahedrite can be seen as a fematinite-based structure that accommodates an ordered arrangement of S-vacancies. 
On one side this sheds new light on the caged nature of tetrahedrite and on its anharmonic structure~\cite{doi:10.1002/adfm.201500766} that are at the origin of the low lattice thermal conductivity. 
On the other side, the S-vacancies are the natural doping mechanism of the Cu-Sb-S compound; indeed, the starting Cu$_3$SbS$_4$ structure is a semiconducting material with a gap of about 0.8~eV and it shows a zT of 0.6 at 600 K under Ge or Sn doping.

We have then investigated the temperature dependence of the electronic transport properties of the compound. For this we have used the Kubo-Greenwood (K-G) formula~\cite{doi:10.1143/JPSJ.12.570,Greenwood_1958} applied to snapshots of first-principles molecular dynamics simulations at different temperatures. 
The use of this approach instead of a Boltzmann transport formalism is here motivated not only by the complex crystal structure of the compound but also by the fact that this method inherently incorporates the interaction between electrons and lattice vibrations as well as, very importantly, anharmonicity, that is supposed to stabilise the high-temperature phase of tetrahedrite for which zero-temperature Density Functional Theory (zero-T DFT) predicts unstable phonons\cite{doi:10.1002/aenm.201200650}.
As we will show, this study reveals the intrinsic transport properties of the compound and their temperature dependence: the Seebeck coefficient is in very good agreement with experimental data and the lattice-limited electrical resistivity shows values that are fairly temperature independent, remarkably low but not too far from the lowest experimental value reported in literature~\cite{Suekuni_2013}.
The Lorenz number turns out to be significantly lower than the free electron value, an important result to quantify lattice and electronic contributions to thermal conductivity.

The structure of the paper is as follows. In the Sec. II, we present the theoretical framework and computational details. The structural characterization of tetrahedrite is presented and discussed in Sec. III. In Sec. IV we analyse the electronic transport properties of the compound.

\section{Theoretical approach and computational details}
The theoretical calculations were performed within DFT in the Kohn-Sham scheme~\cite{dft,KS} (DFT-KS) and using the projector augmented wave method (PAW)~\cite{PhysRevB.59.1758} as implemented in the Vienna ab-initio simulation package (VASP)~\cite{PhysRevB.47.558, PhysRevB.49.14251}. 
We used the Perdew, Burke and Ernzerhof (PBE) exchange-correlation functional~\cite{PhysRevLett.77.3865} for all the calculations and a plane wave energy cutoff of 550~eV; we used a $3\times 3 \times 3$ grid of k-points to sample the Brillouin zone of the conventional unit cell.
Lattice parameters and internal positions were fully relaxed (but constraining the cell to be tetragonal). 
%
%
Formation energies are calculated via the formula $E_{form.}={\begingroup E_0-\sum ^{N_{t}}_{i=1}n_{i}E_{i}\endgroup \over N_{a}}$ 
where $E_0$ is the ground state total energy of the system under consideration that has $N_{a}$ atoms in the unit cell,
$N_{t}$ is the number of different types of atoms in the unit cell, 
and $n_i$ and $E_i$ are the number of atoms of type $i$ and their ground state total energy respectively.

Core binding energies were computed within the $\Delta$ Self-consistent Field ($\Delta$SCF) approximation\cite{jones89,hellman04} at the DFT-KS level of theory. The $\Delta$SCF approximation includes effects from the relaxation of the valence electrons after the core ionization has taken place, reflecting the screening of the core hole created in the photoemission process. Although a core level binding energy is evaluated as the total energy difference between a first core ionized state and the ground state in a spin-polarized calculation, here we follow a spin-paired atomic approximation. To make the unit cell neutral, a compensating electron charge is introduced into the first conduction band. This seems like a good approximation for metals where core-hole screening is efficient. Furthermore, a relativistic Full Potential Linear Muffin-Tin Orbital (FP-LMTO) treatment of the magnetic quantum number\cite{shick96} has been implemented in the QUESTAAL package\cite{questaal,PASHOV2020107065} and used to evaluate the core binding energies of the Cu 2p$_{1/2}$ and 2p$_{3/2}$ electronic states. The smoothed LMTO basis used in this work includes atomic orbitals with $l \le l_{max}=4$ with orbitals 4d of Sb and Cu added in the form of local orbitals.

\footnotetext[1]{Post-processing tool for Kubo-Greenwood electronic transport calculations with VASP code (adapted from Ref.~\onlinecite{CALDERIN2017118}): https://github.com/conodipaola/kg4vasp}
We compute the dynamic Onsager coefficients $\mathcal{L}_{ij}(\omega)$ in the static limit for each ionic configuration in the supercell according to the Kubo-Greenwood (K-G) formula
\cite{MAZEVET201084,CALDERIN2017118,Note1}:
\begin{equation}
\mathcal{L}_{ij}= \lim_{\omega \rightarrow 0} (-1)^{i+j} \frac{2\pi}{3 V} \sum_{n,m,\mathbf{k},\alpha} | \bra{\psi_{n,\mathbf{k}}} \nabla_{\alpha} \ket{\psi_{m,\mathbf{k}}} |^2 (\epsilon_{m,\mathbf{k}}-\mu)^{i-1} (\epsilon_{n,\mathbf{k}}-\mu)^{j-1} \frac{ f(\epsilon_{m,\mathbf{k}})-f(\epsilon_{n,\mathbf{k}}) } { \epsilon_{n,\mathbf{k}}- \epsilon_{m,\mathbf{k}}} \delta(\epsilon_{n,\mathbf{k}} -\epsilon_{m,\mathbf{k}} - \omega)
\label{eq:L}
\end{equation}
where $f(\epsilon)$ is the Fermi-Dirac distribution function, $\mu$ is the chemical potential, $V$ is the volume of the simulation cell, $\alpha=x,y,z$ and $\psi_{l,\mathbf{k}}$ are the KS orbitals (at band $l$ and wave vector $\mathbf{k}$) with corresponding energies $\epsilon_{l,\mathbf{k}}$.
{\color{black} When $\epsilon_{n,\mathbf{k}}- \epsilon_{m,\mathbf{k}}$ goes to zero (the case of intra-band transitions and degeneracies), $[ f(\epsilon_{m,\mathbf{k}})-f(\epsilon_{n,\mathbf{k}}) ] / ( \epsilon_{n,\mathbf{k}} - \epsilon_{m,\mathbf{k}})$ is replaced with $-df(\epsilon_{m,\mathbf{k}})/ d\epsilon_{m,\mathbf{k}} $, as discussed in Ref.~\onlinecite{CALDERIN2017118}.
}
The electrical conductivity is given by
$\sigma=\mathcal{L}_{11}$
while the Seebeck coefficient is
\begin{equation}
S=\frac{\mathcal{L}_{12}}{eT \mathcal{L}_{11} }
\end{equation}
and the electronic contribution to the thermal conductivity is
\begin{equation}
\kappa_e= \frac{1}{e^2T} \big( \mathcal{L}_{22} - \frac{ \mathcal{L}_{12}^2 } {\mathcal{L}_{11} } \big)
\end{equation}
%

The temperature dependence of the transport coefficients is obtained by averaging over snapshots computed via molecular dynamics (MD) simulations~\cite{PhysRevB.85.024102}. We used simulation cells of 232 atoms ($2\times 2 \times 1$ supercells). Initailly we performed NPT runs using the Parrinello-Rahman scheme with implementation of the Nos\'e-Poincar\'e approach for isothermal sampling~\cite{doi:10.1063/1.1416867}; we used a timestep of 1~fs and a period for the thermostat of 1.11~ps; the barostat is used with fictitious masses of $10^{-3}$~amu. These MD simulations were run for 5~ps and this time was enough to calculate the averaged cell parameters at different temperatures.
As a second step, we performed NVT runs using a Langevin thermostat. We used a drift parameter of 1.0 ps$^{-1}$ and performed simulations of about 8.5~ps with a timestep of 1~fs; the first 4.5~ps have been used to safely reach thermal equilibration.
After equilibration we extracted snapshots of nuclear positions every 500 MD steps. For every snapshot we have performed static electronic DFT calculations and used the output to evaluate Eq.~\ref{eq:L}.
While in the {\it ab inito} MD simulations we used a k-points grid of $1\times1\times3$, for every snapshot we used a denser grid of  $3\times3\times6$; for the smearing we used the Fermi-Dirac function with the electronic temperature corresponding to the temperature of the system. The chemical potential obtained from the {\it ab initio} calculation was then consequently used in Eq.~\ref{eq:L}. The Dirac delta functions in Eq.~\ref{eq:L} have been approximated with gaussians of spread 40~meV.
{\color{black} 
The Gaussian broadening is chosen as small as possible in order to remove the small oscillations in the optical conductivity that are due to the discretization of band structure.~\cite{PhysRevE.66.025401}  For instance, at 300K the resistivity changes by less than 5\% when we change the smearing from 20~meV to 70~meV.
}
{\color{black} 
It is also worth to stress that we used a cell of 232 atoms that is tetragonal; the conductivity tensor is however diagonal with the proper symmetry for a cubic system, a quite clear indication of a good convergence of the transport coefficients with respect to the cell size.
}

The electronic transport parameters were also computed using the Boltzmann transport equation within the constant relaxation time approximation (BTE-CRT) using the Wannier interpolation scheme implemented in the BoltzWann code~\cite{PIZZI2014422}. For these calculations we used the DFT band structure of the symmetric high-temperature phase of tetrahedrite.

\section{Crystal structure}
\begin{figure}
\includegraphics[width=6cm]{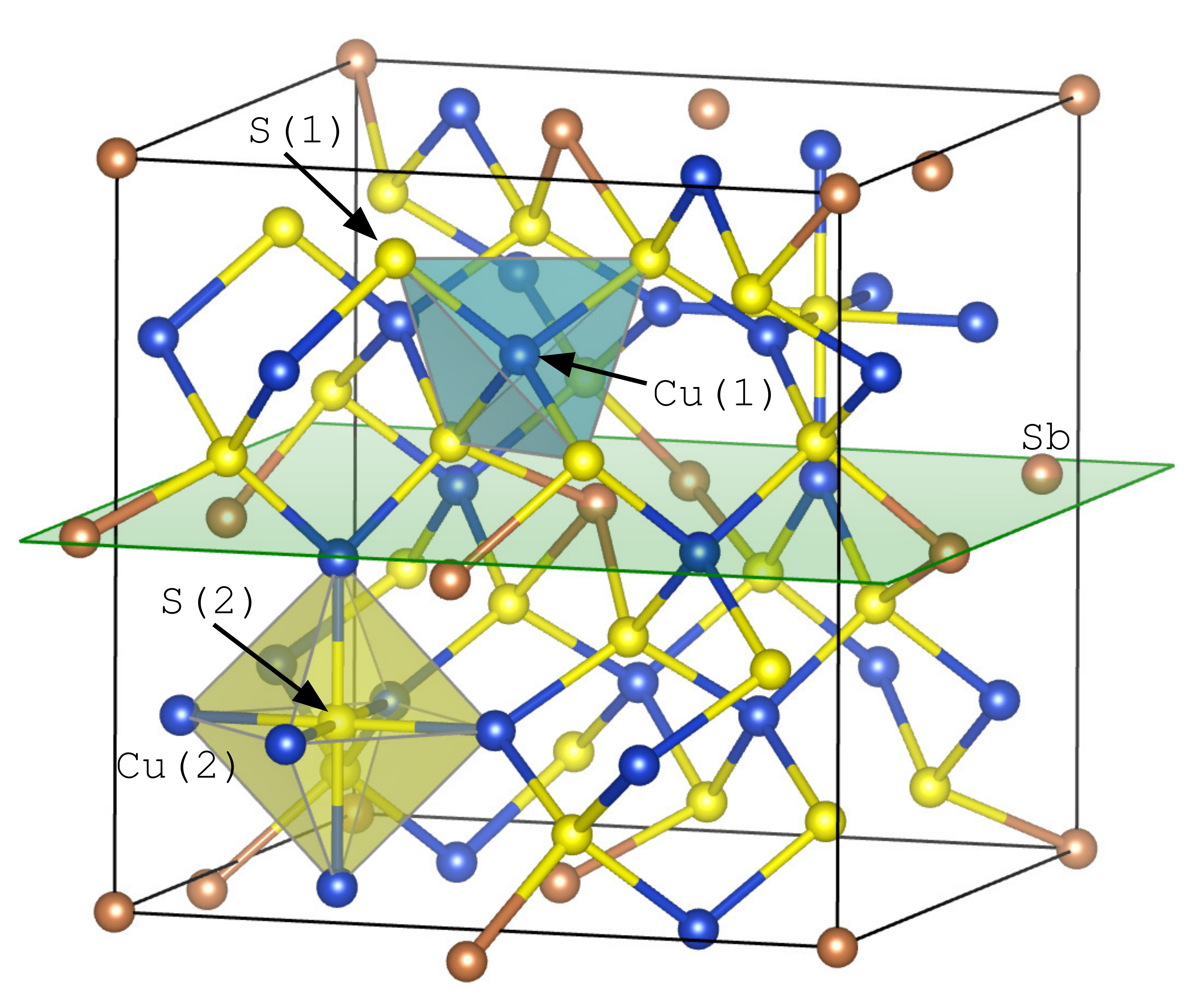} 
\caption{Tetrahedrite cubic conventional cell. S atoms are yellow, Cu atoms are blue and Sb atoms are orange.}
\label{fig:tetrah_perfect}
\end{figure}

Above room temperature tetrahedrite crystallises in a cubic structure ($I\bar{4}3m$). 
The conventional cell of 58 atoms is shown in Fig.~\ref{fig:tetrah_perfect}. 
The experimental lattice parameter of this structure is 10.32~\AA \cite{C9CP00213H,HUANG2018478}. Our DFT calculations predict an equilibrium parameter of 10.40~\AA, in agreement with DFT results reported in literature \cite{doi:10.1002/aenm.201200650}.
In this phase there are 
five distinct crystallographic sites, namely, Cu(1): 12d, Cu(2): 12e, Sb: 8c, S(1): 24g, and S(2): 2a. The Cu(1) atoms are four-fold coordinated as they are at the centre of tetrahedrons of S(1) atoms, whereas the Cu(2) atoms are three-fold coordinated and are located at the vertices of octahedrons at the centre of which there are S(2) atoms. The Sb atoms are on [001] planes at the apex of trigonal pyramids with S(1) atoms.

\begin{figure}
\includegraphics[width=6cm,angle=270,trim={0cm 0cm 0cm 0cm},clip]{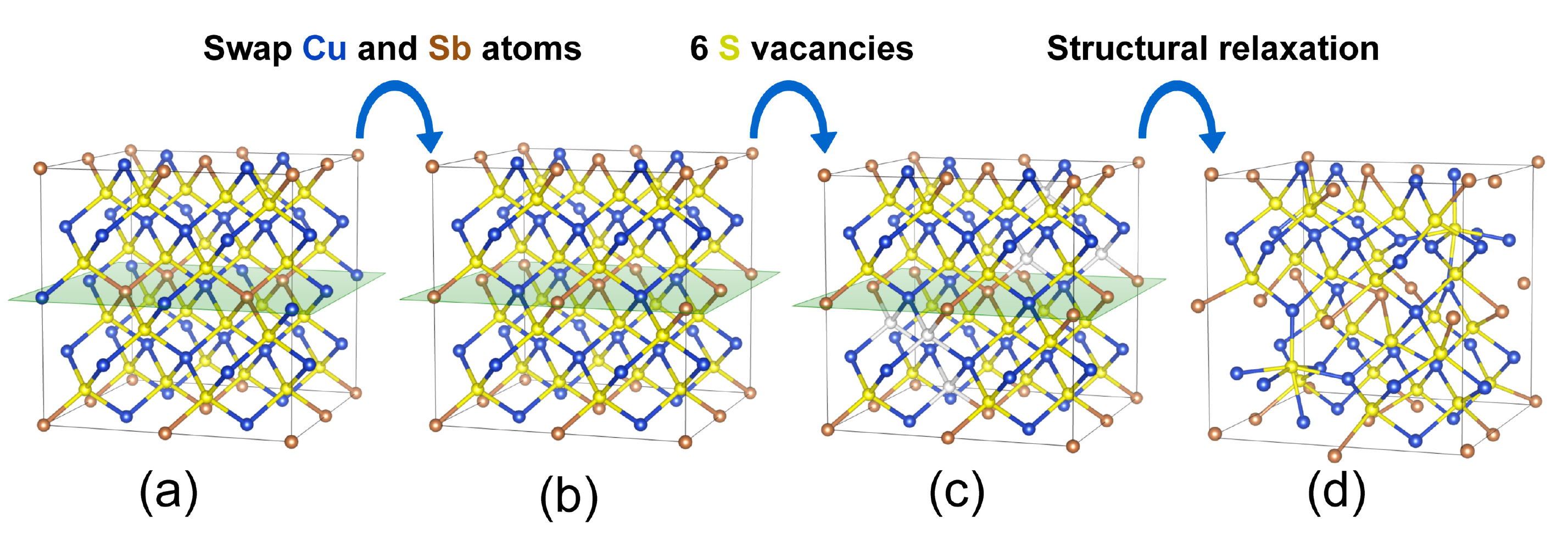}
\caption{(a) $2 \times 2 \times 1$ supercell of fematinite, Cu$_3$SbS$_4$; (b) structure obtained by swapping Cu and Sb atoms within the green $(001)$ plane; (c) structure in which 6 S atoms are removed (white atoms); (d) Relaxed structure obtained starting from (c). Color scheme for the atoms as in Fig.~\ref{fig:tetrah_perfect}.}
\label{fig:panel}
\end{figure}

In order to gain a better understanding of the structure of this complex compound, it is useful to establish a connection between the crystal structure of tetrahedrite and the simpler one of fematinite, a very stable mineral in the Cu-Sb-S system. 
To do this we consider a $2 \times 2 \times 1$ supercell of fematinite structure, containing 64 atoms (Fig.~\ref{fig:panel}a). This supercell is almost cubic as the primitive cell of fematinite is tetragonal ($I\bar{4}2m$) with experimental cell parameters $a=5.39$~\AA ~and $c=10.75$~\AA~\cite{C8TC02481B}; in addition, the average lattice parameter of the supercell is quite close to the one of tetrahedrite.
At this point, in order to establish a structural link between the tetrahedrite and the fematinite structures, one can i) swap Cu and Sb  atoms in the highlighted (001) plane as shown in Fig.~\ref{fig:panel} (b) (or, equivalently, slip the highlighted plane in the $<110>$ direction), and then ii) remove 6 S atoms from the supercell [see Fig.~\ref{fig:panel} (c)], thus recovering the stechiometry of tetrahedrite. 
When the system created in this way is relaxed within zero-T DFT, the resulting structure (Fig.~\ref{fig:panel} (d)) is cubic (with a lattice parameter of 10.39~\AA) and quite similar to the one of tetrahedrite. The only difference is a small distortion of the octahedral structures that is not observed in the high-T phase of tetrahedrite (see Fig.~\ref{fig:tetrah_perfect}). 
This result is not surprising since, in tetrahedrite at low temperatures, there are hints of a structural transition happening around 70~K that modifies the size of the cell and introduces distortions in the local symmetry for $T < 70$ K; this transition is object of an active experimental investigation \cite{Nasonova2016,doi:10.7566/JPSJ.85.014703,PhysRevB.93.064104,Hathwar2019,doi:10.1002/adfm.201909409}. In addition, the instability at low temperatures emerges also from DFT calculations of the phonon spectrum of the high-T symmetric phase \cite{doi:10.1002/aenm.201200650}. 

To conclude the analysis of the link between tetrahedrite and fematinite, it is interesting to compare the stability of the structures suggested in Fig.~\ref{fig:panel}. Our calculations show that the formation energy of the initial and final structures in Fig.~\ref{fig:panel} are the same, -0.241~eV, while the formation energy of the high-T symmetric phase (Fig.~\ref{fig:tetrah_perfect}) is 2~meV higher. It is also interesting to observe that the formation energy of the intermediate step where Cu and Sb are exchanged [the structure of Fig.~\ref{fig:panel} (b)] is only 6 meV higher than the one of fematinite, and that in this structure the clustering of sulphur vacancies in neighbouring positions turns out to be very convenient.
%
%
These results not only strengthen the connection between tetrahedrite and fematinite but also directly show that the Cu-Sb-S crystalline framework can be quite flexible and open to a variety of energetically competitive structures.
%




\section{Electronic structure and XPS}

\begin{figure}
\includegraphics[width=9cm]{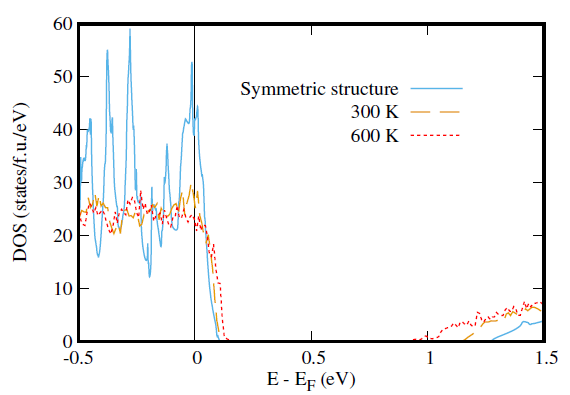}
\caption{{\color{black} Comparison getween the DOS of the symmetric structure in Fig.~\ref{fig:tetrah_perfect} (solid blue line) and the snapshot averaged DOS at 300~K (orange dashed line) and at 600~K (red dotted line). E$_\textrm{F}$ is the Fermi level.
} }
\label{fig:dos}
\end{figure}

{\color{black}
Our calculation for the electronic structure of the symmetric high-temperature phase of tetrahedrite is in agreement with the DFT results reported in literature [see for instance Ref.~\onlinecite{doi:10.1002/aenm.201200650}]. The density of states (DOS) in Fig.~\ref{fig:dos} clearly shows that the compound is metallic but it can be seen as an heavily doped $p$-type semiconductor. Indeed, the Fermi level lies almost at the top of the highest occupied band---the valence band that is mostly formed by sulfur 3p and copper 3d hybridised states; the bottom of the next available empty band---the conduction band---lies at about 1.25~eV from the Fermi level. This band gap is in agreement with other published theoretical results~\cite{doi:10.1002/aenm.201200650}, but it is lower than the experimental gaps reported in literature that vary between 1.7~eV and 1.9~eV.~\cite{doi:10.1021/ja402702x,Jeanloz}
%


The electronic structure of tetrahedrite turns out to be very different from the one of fematinite. Indeed, as discussed in Refs.~\onlinecite{doi:10.1021/acs.jpcc.6b09379, C8TC02481B}, fematinite is a semiconductor with a band gap of around 0.6~eV and the DOS at the top of the valence band is rather smooth and it can be easily fitted with a proper density of states effective mass that can be used in a simple parabolic band model to predict the thermoelectric properties of the compound when $p$-doped. The symmetric phase of tetrahedrite displays instead a rather complex and spiky DOS around the Fermi level, making very difficult the use of simple models to predict the transport properties. 
%


In Fig.~\ref{fig:dos} we also show the effect of thermal motion of atoms on the electronic structure of the compound. For this, we present the DOS of tetrahedrite at 300~K and 600~K, obtained by averaging the DOS of snapshots of the MD simulations. Our results clearly show that thermal motion leads to a strong flattening of the sharp peaks of the DOS of the zero-T symmetric structure as well as to a sensible decrease of the energy gap between the Fermi level and the bottom of the conduction band.

%

}



{\color{black} For this compound it is also important to point out that the chemical composition is often described by} the formula Cu$_{10}^+$Cu$_2^{2+}$Sb$_4^{3-}$S$_{13}^{2-}$~\cite{doi:10.1107/S0108768196014024}, however, the existence of two so different copper atoms (and in different proportion) is still an open question. Indeed, studies of the chemical nature of copper have been carried out by many groups using photoemission spectroscopy (XPS) and X-ray absorption spectroscopy measurements, and some of them report the presence of a Cu$^{2+}$ ions ~\cite{CHETTY2015266, doi:10.1021/acsaem.8b00844}, while others show that both Cu(1) and Cu(2) atoms behave like monovalent ions~\cite{doi:10.7566/JPSJ.85.014703}. 

Our DFT-KS calculations within the $\Delta$SCF scheme show that the two copper atoms in the two inequivalent sites have binding energies of the 2p core states that are very similar and are almost unaffected by the structural distortion discussed in the previous section. In particular, the DFT binding energies of the Cu 2p$_{1/2}$ and 2p$_{3/2}$ states are 951.7 eV and 930.5 eV for the Cu(1) atom and 953.5 eV and 932.3 eV for the Cu(2) atom. 
These values are in very good agreement with the two main peaks at 952~eV and 932~eV observed in XPS experiments. 
These energies are considered the fingerprint of monovalent Cu ions, as they are very similar to what is observed in Cu$_2$O, but they are very different from the spectral peaks in the ranges 940-945~eV and 960-965~eV due to divalent Cu ions observed in CuO. Our results for pristine tetrahedrite in its room temperature phase is therefore compatible with the experimental findings of Tanaka et al.~\cite{doi:10.7566/JPSJ.85.014703} and this suggests that the existence of divalent copper atoms reported in literature might be of extrinsic nature.

\section{Thermoelectric transport coefficients}
%

\begin{figure}
\includegraphics[width=9cm]{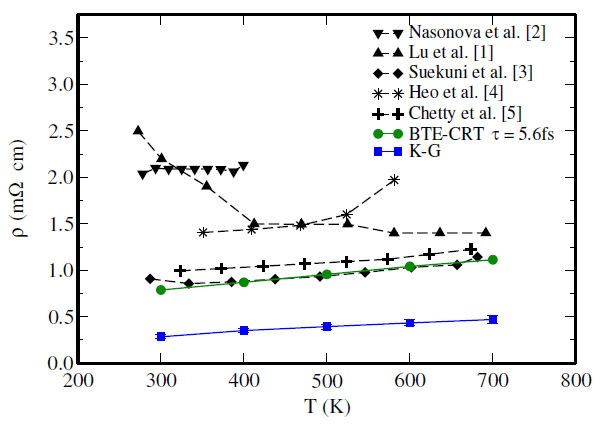}
\caption{Resistivity as a function of temperature. The blue squares (green circles) are the K-G (BTE-CRT) results. The black symbols are the experimental data: the down triangles are from Ref.~\onlinecite{Nasonova2016}, the up triangles from Ref.~\onlinecite{doi:10.1002/aenm.201200650}, the diamonds from Ref.~\onlinecite{Suekuni_2013}, the stars from Ref.~\onlinecite{doi:10.1021/cm404026k}, and the plus symbols from Ref.~\onlinecite{CHETTY2015266}. The error bars in the K-G results are smaller than the symbols used.}
\label{fig:resistivity}
\end{figure}

In this section we discuss the transport properties of the compound and their temperature dependence.
In Fig.~\ref{fig:resistivity} we compare the intrinsic resistivity $\rho=1/\sigma$ of tetrahedrite computed with the K-G approach with the experimental results.
Most of the experimental data show a slow increase of the resistivity above room temperature and this trend is in good agreement with our theoretical results.
The experimental data are distributed on a fairly broad range of values as a result of extrinsic effects (e.g. defects, impurities and polycrystallinity) that in different samples can be of different nature and concentrations. However, our values for the phonon-limited resistivity (about $0.5\times10^{-5}$ $\Omega$ m at 600~K) are not too far from the lowest experimental values reported in literature~\cite{Suekuni_2013,CHETTY2015266}, about $1\times10^{-5}$ $\Omega$ m.
These values are quite low for an undoped mineral in the Cu-Sb-S family.
Indeed, for instance, we recently showed that nano structured Sn doped fematinite at optimal doping for thermoelectric efficiency (a carrier concentration of about $5\times 10^{20}$ cm$^{-3}$ ) displays a resistivity between $1\times10^{-5}$ $\Omega$ m and $1.8\times10^{-5}$ $\Omega$ m between 300K and 600K~\cite{C8TC02481B}.

It is also interesting to observe that the temperature dependence of our theoretical results and of the experimental results with the lowest resistivity~\cite{Suekuni_2013,CHETTY2015266} can also be reproduced fairly well by tuning the carrier lifetime $\tau$ in the BTE-CRT approach that uses the DFT band structure of the symmetric high-temperature phase of tetrahedrite. For instance, as shown in Fig.~\ref{fig:resistivity}, the experimental data in Ref.~\onlinecite{Suekuni_2013} can be fitted in a broad temperature range using a carrier lifetime due to both impurities and phonons of 5.6~fs. 
%
The same fitting can be done on our K-G theoretical results and in this case the carrier lifetime is 16~fs; this is due only to interactions between electrons and lattice vibrations and it quantifies the upper bound of the carrier lifetime in pristine tetrahedrite.

{\color{black}
Quite interestingly, the fact that the temperature dependence of the resistivity is captured by the BTE-CRT approach suggests that the key ingredient to explain this trend is the electronic structure of the compound, while the carrier scattering time due to the electron-phonon coupling seems to depend very weakly on temperature in this material.
}

\begin{figure}
\includegraphics[width=9cm]{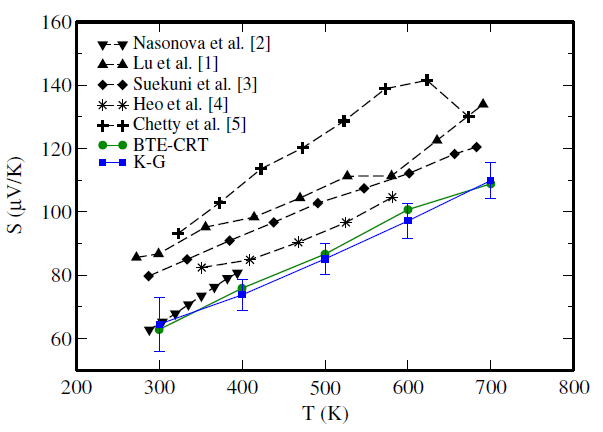}
\caption{Seebeck coefficient as a function of temperature. The symbols are as in Fig.~\ref{fig:resistivity}. }
\label{fig:seebeck1}
\end{figure}

In Fig.~\ref{fig:seebeck1} we show our results for the Seebeck coefficient as a function of temperature. 
The theoretical data are in very good agreement with experiments, and the K-G formalism gives values that are very similar to the ones obtained with the BTE-CRT approach.
The agreement between the exact and the approximate theoretical approaches is not novel (for instance, see the analysis done for simple semiconductors~\cite{PhysRevB.94.085204, PhysRevB.98.201201}) and too surprising because, as we discussed above, a transport coefficient as the resistivity can be
reproduced in a wide range of temperatures using the equilibrium band structure of the symmetric phase and a carrier relaxation time independent of temperature;
the possibility to use a constant relaxation time $\tau$ justifies to a good extent the BTE-CRT approximation in which the Seebeck coefficient (as well as the Lorenz number discussed below) turns out to be independent of $\tau$.

It is important to notice that the values of the Seebeck coefficient are appreciably high in this pristine system.
For instance, the predicted value at 600~K, about 100~$\mu$V/K, is almost as high as the values found in fematinite at optimal Sn doping, about 130~$\mu$V/K~\cite{C8TC02481B}.
These values of the Seebeck coefficient and a low resistivity lead to an undoped compound with a quite high power factor, $S^2/\rho$.
The highest reported experimental values of the power factor at 600~K are between 1.2 and 1.4~mW/(K$^2$m)~\cite{Suekuni_2013,CHETTY2015266,C8CP05479G}, but our calculations suggest the possibility of further improvement. Indeed, the intrinsic upper bound for the power factor predicted here at 600~K is about 2~mW/(K$^2$m).

%

\begin{figure}
\includegraphics[width=9cm]{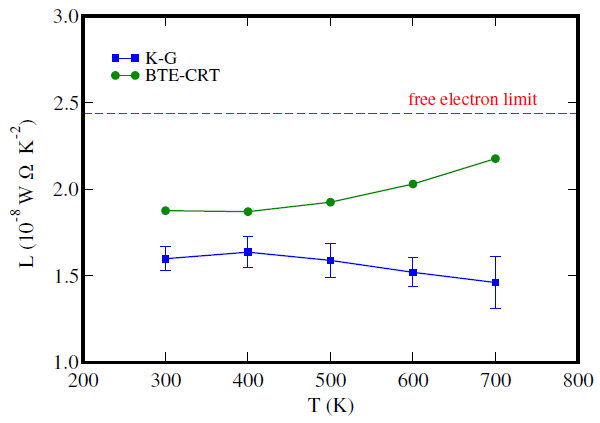}
\caption{Lorenz number as a function of temperature. The blue squares (green circles) are the K-G (BTE-CRT) results; the red dashed line is free-electron value in the Wiedemann-Franz law ($L_0=2.44\times 10^{-8}$W$\Omega$ K$^{-2}$).}
\label{fig:lorenz}
\end{figure}

In Fig.~ \ref{fig:lorenz} we show the results for the Lorenz number, $L=\kappa_e/(\sigma T)$. 
This is a key quantity to estimate the electronic contribution $\kappa_e$ to the thermal conductivity from the experimental values of the electrical conductivity. 
Once $\kappa_e$ is known, it is possible to extract the lattice contribution to the thermal conductivity of the material, $\kappa_L$, from the experimental values ($\kappa_{exp}$), i.e. $\kappa_L=\kappa_{exp}-\kappa_e$. 

Our calculations show that even though, as discussed before, tetrahedrite displays a metallic character above room temperature, the values of $L$ obtained with the K-G formalism are almost temperature-independent but substantially lower than what expected from the free-electron value in the Wiedemann-Franz law ($L_0=2.44\times 10^{-8}$W$\Omega$ K$^{-2}$).
In addition, contrary to what observed for the Seebeck coefficient, the BTE-CRT approach gives values for $L$ that are slightly higher than the K-G ones and with a slightly different temperature trend. Here, the difference between the results from the K-G and BTE-CRT approaches is not surprising because, as shown in other model~\cite{doi:10.1063/1.3656017} and {\it ab initio}~\cite{PhysRevB.94.085204} calculations, the Lorenz number usually tends to be quite sensitive to the details of the carrier scattering mechanisms.

These theoretical predictions for $L$ are interesting and very valuable as they show that the electronic contribution to the thermal conductivity approximated using the free-electron value for $L$ (as done, for instance, in Refs.~\onlinecite{Suekuni_2013, doi:10.1002/aenm.201200650}) can be significantly overestimated, up to about 50$\%$; this thus can lead to a marked underestimation of the intrinsic lattice thermal conductivity of tetrahedrite.
Indeed, for instance, in Ref.~\onlinecite{doi:10.1002/aenm.201200650} the reported value of thermal conductivity is of about 1.45~W/(m K) at 600~K and the lattice contribution estimated using $L_0$ is of about 0.4~W/(m K); if instead the estimation is done using $L$ from the K-G approach, we get a value for $\kappa_L$ of about 0.8~W/(m K) ($\kappa_L\approx 0.6$~W/(m K) using $L$ from the BTE-CRT approach), suggesting that in experiments the lattice thermal conductivity of the compound is certainly low, but it can be very similar to the electronic contribution.

\section{Conclusions}
We have presented a detailed {\it ab initio} study of the structural and electronic transport properties of the high-temperature phase of tetrahedrite.
We have shown that the structural variety of the Cu-Sb-S network allows a description of tetrahedrite in terms of a fematinite-like crystal modified by an ordered arrangement of S-vacancies.
%
%
While this structure presents two crystallographically inequivalent copper atoms, our calculations have shown that these atoms experience similar local chemical environments. Indeed, they have very similar core level binding energies and this helps the analysis of XPS measurements in this compound.
The Kubo-Greenwood approach has allowed us to predict two important thermoelectric quantities, the phonon-limited electrical resistivity and the Lorenz number.
The predicted resistivity turns out to be quite low for an undoped compound in the Cu-Sb-S system. 
The lowest experimental data reported in literature are not too far from the predicted intrinsic values: 
this clearly shows that the quality of the samples studied in experiments is high, but it also suggests the possibility of further improvements of the electronic transport properties.
The Lorenz number has turned out to be substantially lower than what expected for the free-electron value, often used to estimate the electronic and lattice contributions to the thermal conductivity in experiments. Thus our result provides a more accurate reference to analyse thermal transport in this compound.
Finally, it is important to stress that the K-G approach has been key to predict transport properties in this system. 
Indeed, the zero-temperature DFT phonon calculations for the high temperature phase result in imaginary frequencies and lattice instabilities, which make hard to apply the exact BTE approach that includes electron-phonon coupling.
Our analysis has also allowed us to show that the use of a less computationally demanding BTE-CRT approach is quite effective to reproduce the temperature trends of transport quantities such as the resistivity and the Seebeck coefficient in this complex compound. 

\section{Acknowledgements}
This work was supported by the UK Engineering and Physical Sciences Research Council (EPSRC) (Grant EP/N02396X/1). This article was also developed based upon funding from the National Renewable Energy Laboratory, managed and operated by the Alliance for Sustainable Energy, LLC for the U.S. Department of Energy (DOE) under contract no. DE-AC36-08GO28308. N.B. and C.W. acknowledge the ARCHER UK National Supercomputing Service and the UK Materials and Molecular Modelling Hub for computational resources, which are partially funded by EPSRC (EP/P020194/1). N.B. also acknowledges the Cirrus UK National Tier-2 HPC Service at EPCC. C.W and N.B acknowledge Dr Glenn Jones (Johnson \& Matthey, Sonning) for support and helpful discussions. C.W. gratefully acknowledges the support of NVIDIA Corporation with the donation of the Tesla K40 GPUs used in this research.

\bibliography{biblio}
\end{document}